\documentclass[12pt]{article}

\usepackage[cp1251]{inputenc}

\input{epsf}

\usepackage[pdftex]{graphicx}
\begin{document}

%\author{S.~V.~Poltavtsev}
%\author{I.~I.~Ryzhov}

%\author{R.~V.~Cherbunin}
%\author{A.~V.~Mikhailov}

%\author{N.~E.~Kopteva}
\date{}
\author{  G.~G.~Kozlov, I.~I.~Ryzhov, V.~S.~Zapasskii}

%\affiliation{Spin-Optics laboratory, St.~Petersburg State University, 198504 St.~Petersburg, %Russia}  

%\author{K.~V.~Kavokin}
%\affiliation{Spin-Optics laboratory, St.~Petersburg State University, 198504 St.~Petersburg, Russia}
%\affiliation{A.~F. Ioffe Physical-Technical Institute, Russian Academy of Sciences, 194021 %St.~Petersburg, Russia}

%\author{V.~S.~Zapasskii}
%\affiliation{Spin-Optics laboratory, St.~Petersburg State University, 198504 St.~Petersburg, %Russia}  

%\author{I.~V.~Ignatiev}
%\affiliation{Spin-Optics laboratory, St.~Petersburg State University, 198504 St.~Petersburg, %Russia}

%\author{P.~V.~Lagoudakis}
%\affiliation{Department of Physics \& Astronomy, University of Southampton, Southampton SO17 %1BJ, United Kingdom}

%\author{A.~V.~Kavokin}
%\affiliation{Spin-Optics laboratory, St.~Petersburg State University, 198504 St.~Petersburg, %Russia}
%\affiliation{Department of Physics \& Astronomy, University of Southampton, Southampton SO17 %1BJ, United Kingdom}

%\date{\today}
%\maketitle 

\title{Light scattering in the medium with fluctuating gyrotropy: application to spin noise spectroscopy}
\maketitle
\begin{abstract}
The spin noise signal in the Faraday-rotation-based detection technique can be considered equally correctly either as a manifestation of the spin-flip Raman effect or as a result of light scattering in the medium with fluctuating gyrotropy. In this paper, we present rigorous description of the signal formation process upon heterodyning of the field scattered due to fluctuating gyrotropy.  Along with conventional single-beam experimental arrangement, we consider here a more complicated, but more informative, two-beam configuration that implies the use of an auxiliary light beam passing through the same scattering volume and delivering additional scattered field to the detector. We show that the signal in the spin noise spectroscopy arising due to heterodyning of the scattered field  is formed only by the scattered field components whose wave vectors coincide with those of the probe beam. Therefore, in principle, the detected signal in spin noise spectroscopy can be increased by increasing overlap of the two fields in the momentum space.  We also show that, in the two-beam geometry, contribution of the auxiliary (tilted) beam to the detected signal is represented by Fourier transform of the gyrotropy relief at the difference of two wave vectors. This effect can be used to study spin correlations by means of noise spectroscopy.  
\end{abstract}

\section*{Introduction}

The spin noise spectroscopy (SNS), first realized in \cite{Zap}, has turned nowadays into a powerful method of studying magnetic resonance and spin dynamics in atomic and semiconductor systems (see, e.g., \cite{Zap1, Oest2, Sin} . The most fascinating results of application of the SNS with the greatest progress in sensitivity of the measurements were achieved in physics of semiconductor structures, where the novel technique has allowed one not only to considerably move ahead in the magnetic resonance spectroscopy, but also to discover fundamentally new opportunities of research.  Specifically, it has been established that optical spectroscopy of spin noise (that implies measuring wavelength dependence of the spin noise power) makes it possible to decipher inner structure of optical transitions \cite{Zap2}. Correlation nature of the SNS allowed one to realize, on its basis, a sort of pump-probe spectroscopy \cite{TC}. Effective dependence of the spin-noise signal on the light-power density (on the beam cross section) was used to demonstrate the SNS-based 3D tomography \cite{To, To1}.  Due to high sensitivity of the SNS, it appeared possible to detect magnetic resonance of quasi-free carriers in a single quantum well 20 nm thick \cite{Glazov}, to observe the spin-noise spectrum of a single hole spin in a quantum dot \cite{singlehole}, and to realize  magnetometry of local magnetic fields (including field of polarized nuclei) in a semiconductor \cite{R1,R}.  Due to these remarkable capabilities of the new technique, it acquired a great popularity during the last decade. 

At the same time, fundamental mechanism underlying the effect of magnetic resonance in the Faraday rotation noise spectrum remains so far, to a considerable extent, unexplored. Theoretically, it has been shown in 1983 \cite{Gorb} that this effect is closely related to the spin-flip Raman scattering, and the detected signal of magnetic resonance is the result of heterodyning of the scattered light (with shifted frequency), with the local oscillator provided by the probe laser beam. In this case, the standard experimental geometry we use in the conventional SNS may appear to be far from optimal. Indeed, we usually collect, on the photodetector, only the scattered light lying within the solid angle of the probe beam, whereas indicatrix of the Raman-scattered light may be fairly isotropic. It means that, in the standard experimental geometry, most part of the scattered light is lost. Therefore, it looks like the detected signal, in the SNS, can be considerably increased by collecting more efficiently the scattered light. Still, even if this simple picture is correct, it is not easy to correctly design the experimental setup to take advantage of the additional scattered field in full measure. First experiments carried out in this direction \cite{Scal} and our preliminary analysis of the problem have shown that favourable solution of this experimental task can be achieved only with allowance for  all the factors affecting the heterodyning process (wave fronts of the reference and scattered waves, shape of the beam, volume of the scattering medium, shape and dimensions of the photosensitive  surface, correlation properties of the gyrotropy, etc. ). Actually, this problem, which we consider to be fundamental for the SNS method, is rather complicated and needs to be  analyzed carefully and rigorously, with the results of the treatment applicable to real experimental conditions. In our opinion, computational details of such a treatment and prticularities of used apprpximations are also highly important. 

In this paper, we present such a treatment for a focused Gaussian probe beam propagating through the medium with fluctuating gyrotropy and analyze in detail mechanism of the intensity-noise signal formation  due to heterodyning of the scattered field on the detector. We also propose a two-beam  experimental arrangement, with the auxiliary light beam tilted with respect to the probe, that makes it possible to get information about the spatiotemporal correlation function of gyrotropy of the studied system (remind that in conventional SNS only spatially averaged temporal correlation function is revealed).

The paper is organized as follows. In Section 1, for completeness of the narrative, we present a brief explanation of what is the Gaussian beam and introduce a model of the polarimetric detector used in our further analysis. We show here that the detected signal in SNS is contributed only by the scattered field that, in the momentu, space, coincides with that of the probe. In Section 2, we present basics of the single-scattering theory, apply it to the medium with gyrotropy randomly modulated in space, and calculate the observed polarimetric signal. In Sections 3 and 4, we calculate the noise signal observed in the two-beam configuration, when the auxiliary beam propagating though the medium at some angle to the main probe beam does not hit the detector and contributes to the signal only by its scattered field. We show that the spin-noise signal, under these conditions, is proportional to the Fourier component of the spatial correlation function of gyrotropy at spatial frequency equal to difference between the two wave vectors. In Section 5, we present calculations for the model of independent paramagnetic particles (spins) and show that the signal produced by the auxiliary tilted beam is of the same order of magnitude as the one produced by the main probe and, hence, can be easily detected using the same experimental setup. 

\section{Detecting polarimetric signal in a confined laser beam}

In  the simplest version of the light-scattering problem, the probe beam can be taken in the form of a plane wave. However, in the SNS experiments under consideration, when two light beams are supposed to be used, with their spatial localization being of crucial importance, this approximation proves to be inappropriate.  So, we will treat Gaussian beams whose electric  fields ${\bf E_p(r)}$ are defined by the expression
    \begin{equation}
               {\bf E_p(r)}=e^{\imath (kZ-\omega t)}  
               kQ\sqrt{8 W\over c}
               {(\cos\eta,0,-\sin\eta)\over (2k+\imath Q^2Z)}
               \exp\bigg [-{kQ^2(X^2+y^2)\over 2(2k+\imath Q^2Z)}\bigg ]
               \label{a0}
               \end{equation}
               where ${\bf r}=(x,y,z)$, $k\equiv \omega/c$ ($\omega$ is the optical frequency and $c$ is the speed of light), $W$ -- beam intensity, and 
                           $$
                           \left(\matrix{X\cr Z}\right )=\left(\matrix{\cos\eta &-\sin\eta \cr \sin\eta & \cos\eta}\right )\left(\matrix{x\cr z}\right )
                           $$
    Field (\ref{a0})  satisfies Maxwell's equations and represents the  beam propagating  in the $zx$ plane at the angle $\eta$ with respect to $z$ axis ($\eta$ is assumed to be snall) and polarized mostly in $x$ -direction.  The parameter $Q$ defines the $e$-level half width $2w$ of the beam waist  by relationship $w=1/Q$.  $w$ should be greater than  the wavelength $\lambda= 2\pi c/\omega$.  In our estimations,  we accept $\lambda\sim 1\ \mu$m and $w\sim 30\ \mu$m.

In the SNS experiments, we detect small fluctuations of the optical field polarization, and, therefore, to calculate correctly the SNS signal, we have to specify the model of  polarimetric detector. We suppose the detector  to be  comprised of two photodiodes PD1 and PD2 (Fig.\ref{fig1}) arranged in  two arms of the polarization beamsplitter (BS).
 The  output signal  $U$ is obtained by subtracting photocurrents of the two photodiodes and (to within some unimportant factors) are given by the expression
\begin{equation}
  U={\omega\over2\pi}\int _{0}^{2\pi/\omega} dt\int_{-l_x}^{l_x}dx\int_{-l_y}^{l_y}dy\bigg [
  \hbox{Re }^2E_x(x,y,L)-\hbox{Re }^2E_y(x,y,L)
   \bigg ],
   \label{a1}
  \end{equation}
where $E_{x,y}$ are the $x$ and $y$ components of the complex input optical field $\bf E$, $2l_{x,y}$ are the dimensions of sensitive areas of the photodiodes along the $x$ and $y$ directions.  We ascribe  physical sense to  real part of the complex optical field and,  as seen from Eq. (\ref{a1}), the output  signal $U$ represents the difference between intensities of  the input optical field in the $x$ and $y$ polarizations integrated over sensitive areas of the photodiodes and averaged over the optical period $2\pi/\omega$.  
   
In our case, the input optical field $\bf E$ can be presented as a sum of   the probe field  $\bf E_0$ (Re $\bf E_0={\cal E}_0$) and the field $\bf E_1$ (Re $\bf E_1={\cal E}_1$)  arising  due to    scattering of the probe  beam by   the sample with spatially fluctuating gyrotropy. Then, the first-order (with respect to $\bf E_1$) contribution $u_1$ to   the polarimetric signal  can be written as
 \begin{equation}
 u_1={\omega\over\pi}\int _{0}^{2\pi/\omega} dt\int_{-l_x}^{l_x}dx\int_{-l_y}^{l_y}dy 
\bigg [
{\cal E}_{x0}(x,y,L){\cal E}_{x1}(x,y,L)-{\cal E}_{y0}(x,y,L){\cal E}_{y1}(x,y,L)\bigg ]
\label {a2} 
\end{equation}

This formula shows that the observed signal can be thought of as a result of heterodyning (mixing) of the unperturbed probe field ${\cal E}_0$ with the field of scattering ${\cal E}_1$. 
 Equation (\ref{a2}) also shows that , for sufficiently large dimensions of the detector ($l_{x,y}\gg\lambda=2\pi/k$),  polarimetric signal $u_1$ represents {\it projection of the scattered field} (in the momentum space) onto the field of the probe beam. This means, in turn, that  this signal is controlled by the fraction of the scattered field whose distribution in space, to a certain extent, reproduces the field of the probe beam.  Specifically, when the probe field represents a plane wave ${\bf E}_0\sim e ^{\imath {\bf q}_0{\bf r}}$ with the wave vector ${\bf q}_0$, and the scattered field can be presented by a superposition of the plane waves ${\bf E}_1\sim \int d{\bf q}e^{\imath {\bf qr}}{\bf S (q)}$, the signal $u_1$ appears to be proportional to the component of the scattered field at the spatial frequency ${\bf q}_0$: $u_1\sim {\bf S(q}_0)$.  

Let us now calculate the scattered field ${\bf{\cal E}} _1$.

\section{ Polarimetric signal in the medium with fluctuating gyrotropy}

In this section, we consider scattering of a monochromatic light beam by the medium with randomly inhomogeneous (spatially fluctuating) gyrotropy. In this case, polarization of the medium ${\bf P(r) }$ can be expressed through the  electric field  ${\bf E(r)}$ by the expression
\begin{equation}
{\bf P(r)}=\imath [{\bf E(r)G(r)}]=\imath {\bf E(r)\times G(r)}
\label{a3}
\end{equation}
where ${\bf G(r)}$ is the spatially dependent gyration vector. At this stage of our treatment, we assume the gyration vector to be time-independent. Then,   Maxwell's equations for  the electromagnetic field  in the medium can be reduced to the form:
\begin{equation}
\Delta {\bf E}+k^2{\bf E}=-4\pi k^2{\bf P}-4\pi \hbox{ grad div }{\bf P}, \hskip10mm k\equiv {\omega\over c}
\label{a4}
\end{equation}
We will search for solution of this equation in the form of series in powers of ${\bf G(r)}$. The zero order term ${\bf E_0(r)}$  represents the probe beam field  which we consider to be known. The first order term ${\bf E_1(r)}$ corresponds to  the single-scattering approximation which is sufficient for our consideration.  
This term  satisfies  the equation
\begin{equation}
\Delta {\bf E_1}+k^2{\bf E_1}=-4\pi\imath k^2{\bf E_0(r)\times G(r)}-4\pi\imath \hbox{ grad div }{\bf E_0(r)\times G(r)}
\label{a5}
\end{equation}

Solution of this equation can be expressed in terms of  Green’s 
function $\Gamma({\bf r})=-\exp (\imath kr)/4\pi r$ of the Helmholtz equation $[\Delta+k^2]\Gamma ({\bf r})=\delta({\bf r})$:
\begin{equation}
     {\bf E_1(r)}=\imath\int  {\exp (\imath k|{\bf r-r'}|)\over {\bf |r-r'|}}\bigg [ k^2{\bf E_0(r')\times G(r')}+ \hbox{ grad div }{\bf E_0(r')\times G(r')}\bigg ]d^3{\bf r'}
      \label{a6}
      \end{equation}
 Let the sample (we call  “sample”  the region where ${\bf G(r)}$ is  nonzero)  be placed in the vicinity of the origin of our coordinate system $x,y,z$.  Let   the photosensitive surface of the polarimetric detector be parallel to the $xy$ plane and the detector itself be set at $z=L$, with $L$ being large  compared with the sample dimensions.  Then,  as seen from Eq. (\ref{a6}),  the scattered field can be presented as a sum  of two contributions:   
\begin{equation}
{\bf E_1(r)=E_{1}^1(r)+E_{1}^2(r)}
\label{a7}
\end{equation}
$$
{\bf E_{1}^1(r)}\equiv {\imath k^2\over L}\int  \exp (\imath k|{\bf r-r'}|){\bf E_0(r')\times G(r')} d^3{\bf r'}
$$
$$
{\bf E_{1}^2(r)}\equiv {\imath\over L}\int  \exp (\imath k|{\bf r-r'}|)\hbox{ grad div }{\bf E_0(r')\times G(r')}d^3{\bf r'}={1\over k^2}\hbox{ grad div }{\bf E_1^1(r)}
$$

We will concentrate on calculating  the part ${\bf E_1^1(r)}$  of the scattered field  because, in what follows, we will  need this field at small scattering angles  and, in this case, as it can be directly checked, only ${\bf E_1^1(r)}$ is of importance.
           
 We take the probe beam in the form of Eq. (\ref{a0}) at $\eta =0$, with the angle $\phi$ specifying beam polarization in the $xy$ plane.  
%\newpage 
Then,  the probe field acquires the form 
  \begin{equation}
  {\bf E_0(r)}=e^{\imath [kz-\omega t]}  
   kQ\sqrt{8 W\over c} {(\cos\phi,\sin\phi,0)\over (2k+\imath Q^2z)}
      \exp\bigg [-{kQ^2(x^2+y^2)\over 2(2k+\imath Q^2z)}\bigg ], \hskip10mm {\bf r}=(x,y,z)
                         \label{a8}
                         \end{equation}
We need  this field in two substantially separated spatial regions:   firstly,  in Eq.(\ref{a2}) at large  values of $z\sim L$ and,     secondly, in Eq.(\ref{a7}) at relatively small values of $z$  within the sample. Calculation for $z\sim L$  shows that the field ${\bf {\cal E}}_0$ entering Eq.(\ref{a2}) has the form
     
 \begin{equation}
               \left (\matrix {{\cal E}_{x0}(x,y,L)\cr {\cal E}_{y0}(x,y,L)}\right )=
\left (\matrix {\cos\phi\cr\sin\phi}\right ) \sqrt{8 W\over c}
             {k\over  QL}\hskip1mm \sin \bigg [ kL-\omega t
             +{k[x^2+y^2]\over 2L}
             \bigg ]  
                                               \exp\bigg [-{k^2(x^2+y^2)\over  Q^2L^2}\bigg ]
                                               \label{a9}
                                               \end{equation}

While deriving these expressions, we assumed that $L> z_c =4\pi w^2/\lambda$ ($z_c$  is
the Rayleigh length). To calculate the scattered field by Eq. (\ref{a7}), one needs the 
field (\ref{a8}) at $z<z_c$. In this limit, Eq. (\ref{a8}) can be 
simplified:

\begin{equation}
                         {\bf E_0(r)}=e^{\imath [kz-\omega t]}  
                         Q\sqrt{8 W\over c}
                         {(\cos\phi,\sin\phi,0)\over 2}
                         \exp\bigg [-{Q^2(x^2+y^2)\over 4}\bigg ],\hskip10mm z<z_c
                         \label{a10}
                         \end{equation}
Using this relationship, one can calculate the scattered field 
${\bf E_1^1(r)}$ (\ref{a7}) and obtain, for  real parts of ${\cal E}_{x1}$ and ${\cal E}_{y1}$  
entering Eq. (\ref{a2}), the following expressions:
 
 \begin{equation}
\left (\matrix{{\cal E}_{x1} \cr
{\cal E}_{y1}}\right )
=\left (\matrix{-\sin\phi\cr\cos\phi}\right ) \sqrt{2 W\over c}{Q k^2\over L}
 \int   \sin [k|{\bf r-r'}|+kz'-\omega t]                           
                         \exp\bigg [-{Q^2(x'^2+y'^2)\over 4}\bigg ]
 G_z({\bf r'}) d^3{\bf r'}
\label{a11}
\end{equation}

Using Eq.  (\ref{a2}) and explicit expressions (\ref{a9}) and (\ref{a11}) for the probe ${\cal E}_0$ and scattered ${\cal E}_1$ fields, we can calculate the polarimetric signal. While averaging  the product 
 ${\cal E}_{x0}{\cal E}_{x1}$ over the optical period, we come to the integral
    $$
  {\omega\over \pi}\int_0^{2\pi/\omega}{\cal E}_{0x}{\cal E}_{1x}\hskip1mm dt\sim
 {\omega\over \pi}\int_0^{2\pi/\omega}\sin [k|{\bf r-r'}|+kz'-\omega t]\hskip1mm
 \sin \bigg [ kL-\omega t
              +{k[x^2+y^2]\over 2L}
              \bigg ]  \hskip1mm dt =
  $$
$$
=\cos \hskip1mm k\bigg [
z'+|{\bf r-r'}|-L-{x^2+y^2\over 2L}
\bigg ]
$$
  The same is obtained for  ${\cal E}_{y0}{\cal E}_{y1}$. Now,  Eq. (\ref{a2}) gives

 \begin{equation}
    u_1= -{4Wk^3 \sin[2\phi]\over cL^2}                     
                     \int_{-l_x}^{l_x}dx\int_{-l_y}^{l_y}dy 
                                          \exp\bigg [-{k^2(x^2+y^2)\over  Q^2L^2}\bigg ]
                                          \times
           \label {a12} 
\end{equation}
$$
\times \int  
\hskip1mm
\cos \hskip1mm k\bigg [
z'+|{\bf r-r'}|-L-{x^2+y^2\over 2L}
\bigg ]                         
      \hskip1mm \exp\bigg [-{Q^2(x'^2+y'^2)\over 4}\bigg ]
          G_z({\bf r'}) d^3{\bf r'},
$$
with  ${\bf r}=(x,y,L)$ and ${\bf r'}=(x',y',z')$. The external integration over $dxdy$ runs over the  detector sensitive area, and, therefore, $|x|,|y|<l_{x,y}\ll L$. We assume that  dimensions of the detector $l_{x,y}$  
 exceed the size $L\lambda/2\pi w$ of the probe beam spot at the detector 
  (see Eq. (\ref{a9})).  Then, $x$ and $y$ can be estimated as $x,y\sim L\lambda/2\pi w$.
  The internal integration $d{\bf r'}$ runs over the 
 irradiated volume of the sample. For this reason $x',y'\sim w$ and $z'$ is of the order of the sample length $l_s$.
 Taking into account that $L\lambda/2\pi w,w,l_s\ll L$, we obtain 
  the following expansion for the factor $|{\bf r-r'}|$:
 
   \begin{equation}
   |{\bf r-r'}|\approx L+{x^2+y^2\over 2L}+
    {x'^2+y'^2\over 2L}-
   {xx'+yy'\over L}-{z'}.
  \label{a13}
    \end{equation}
Note that the term $\sim z'^2$ vanishes. Further estimates show that the term $(x'^2+y'^2)/2L$
 can be omitted because in our case $k(x'^2+y'^2)/2L<\pi/4$ and, finally, we have
 \begin{equation}
   |{\bf r-r'}|\approx
 L+{x^2+y^2\over 2L}
    -{z'}
   -{xx'+yy'\over L}
  \label{a14}
   \end{equation} 
 
Using this formula, we can evaluate  the product of the cosine functions in (\ref{a12}) as

   \begin{equation}
     \cos \hskip1mm k
   \bigg [
   z'+|{\bf r-r'}|-L-{x^2+y^2\over 2L}
   \bigg ]  =
      \cos  k\bigg [
             {xx'+yy'\over L}
            \bigg ]  
   \label{a15}
   \end{equation}

 As was mentioned above, the detector dimensions are assumed to be  greater than the size of the probe beam spot: 
$l_{x,y}> L\lambda/2\pi w$.
 This allows one to extend  integration over  the detector surface in (\ref{a12}) to infinity: $|l_{x,y}|\rightarrow\infty$
 and to calculate all  integrals using the    formula   
      \begin{equation}
       \int dx \exp [-\alpha x^2+\imath\beta x ]=\sqrt{\pi\over \alpha}\exp \bigg (-{\beta^2\over 4\alpha}\bigg ).
       \label{a16}
       \end{equation}
  For example, the integral with  cosine function in  Eq.(\ref{a15}) (we denote it $I_1$) can be calculated as follows:

   \begin{equation}
   I_1=\int_{-\infty}^\infty dx \int_{-\infty}^\infty dy\hskip1mm  \exp\bigg [-{k^2(x^2+y^2)\over  Q^2L^2}\bigg ] \hskip1mm \cos  k\bigg [
             {xx'+yy'\over L}
            \bigg ]=
\label{a17}   
\end{equation}
   $$
   =\hbox{ Re }
   \int_{-\infty}^\infty dx \int_{-\infty}^\infty dy\hskip1mm  \exp\bigg [-{k^2(x^2+y^2)\over  Q^2L^2}+\imath k 
                   {xx'+yy'\over L}
                     \bigg ]= 
  $$             
   $$
      =\hbox{ Re }
      \int_{-\infty}^\infty dx 
       \exp\bigg [-{k^2 x^2\over  Q^2L^2}+\imath k 
                            {xx'\over L}
                              \bigg ]
            \hskip2mm\int_{-\infty}^\infty dy\hskip1mm  \exp\bigg [-{k^2y^2\over  Q^2L^2}+\imath k 
                      {yy'\over L}
                     \bigg ]=
     $$  
     $$
     ={\pi Q^2L^2\over k^2} \exp  \bigg (
                -{[x'^2+y'^2]Q^2\over 4}
                \bigg )  
         $$                  
      
Substituting (\ref{a17}) into (\ref{a12}), we obtain the following expression for the  polarimetric signal:

            \begin{equation}
                  u_1= -{4Wk \pi Q^2\sin[2\phi]\over c}                     
               \int_V   
                   \exp\bigg [-{Q^2(x'^2+y'^2)\over 2}\bigg ]
                            G_z({\bf r'}) d^3{\bf r'}                   
                         \label {a18} 
              \end{equation}

Remind that this formula is valid if the sample length $l_s$ is smaller than the Rayleigh length, $l_s<z_c$ (see definition of the Rayleigh length after Eq. (\ref{a9})) and
 the probe beam spot is smaller than the detector photosensitive area, $l_{x,y}\gg L\lambda/2\pi w$. It is
 seen from Eq. (\ref{a18}) that the  polarimetric signal is, in fact, proportional to $z$-component of the gyration
 averaged over  irradiated volume of the sample, as is usually implied intuitively.

Equation (\ref{a18}) allows  one to obtain the expression for the magnetization noise power spectrum observed in the SNS.  In this case, ${\bf G(r)}$ is  proportional to instantaneous spontaneous magnetization of the sample randomly fluctuating both in space, and in time. If characteristic  frequencies of this field are much lower than the optical frequency $\omega$,  one can use Eq. (\ref{a18}) for calculating the random polarimetric signal by substituting ${\bf G(r)}\rightarrow {\bf G(r,}t)$.
 The noise power spectrum ${\cal N}(\nu)$ is defined as Fourier transform of correlation function of the polarimetric signal. 
 Using Eq. (\ref{a18}), the noise power spectrum ${\cal N}(\nu)$ can be expressed in terms  of the  spatiotemporal correlation function of the gyrotropy ${\bf G(r,}t)$:
\begin{equation}
{\cal N}(\nu)=\int dt\langle u_1(t)u_1(0)\rangle e^{\imath \nu t}={16 W^2k^2\pi^2Q^4\sin^2[2\phi]\over c^2}\times
\label{a19}
\end{equation}
$$
\times
\int dt\hskip1mm  e^{\imath \nu t}\int_V d^3{\bf r}\int_V d^3{\bf r'}
\exp\bigg [-{Q^2(x'^2+y'^2+x^2+y^2)\over 2}\bigg ]
                                     \langle G_z({\bf r'},0)G_z({\bf r},t)\rangle
  $$

\begin{figure}
    \includegraphics[width=.8\columnwidth,clip]{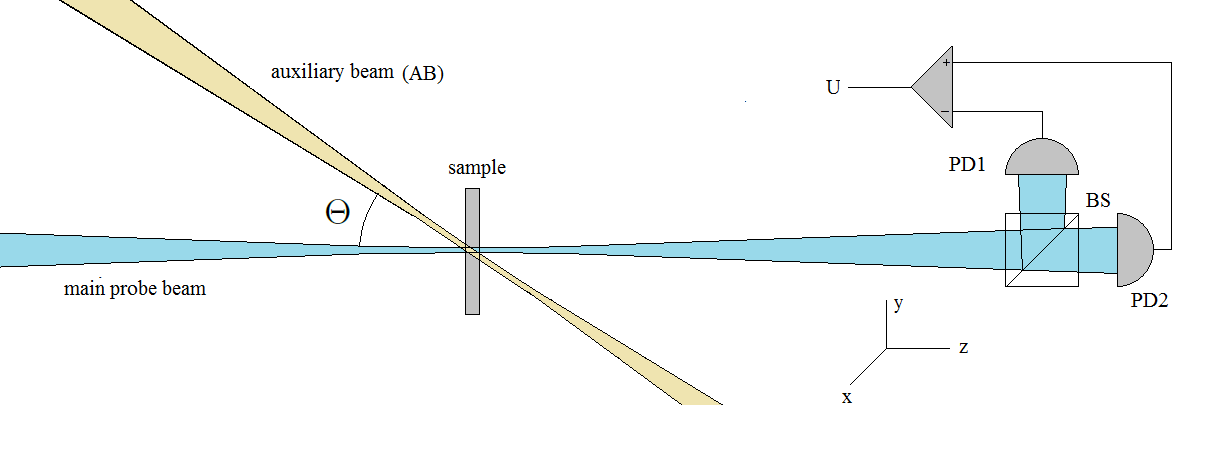}
     %\epsffile{fig1.png}
     \caption{Detecting the noise signal produced by the auxiliary beam. }
     \label{fig1}
     \end{figure}

  To calculate the correlation function $ \langle G_z({\bf r'},0)G_z({\bf r},t)\rangle$ entering Eq. (\ref{a19}), one should specify a particular model of the gyratropic medium. The example of such a model (the   model  of independent  paramagnetic atoms with fluctuating magnetization) will be described in Section 5.      In the next section, we will calculate  the plarimetric signal produced by an auxiliary tilted beam that produces a scattered field but does not irradiate  the detector (see Fig. \ref{fig1}). 
   
\section{Detecting scattered field of a tilted beam }

Let the sample be illuminated by an auxiliary light beam (AB) propagating at the angle $\Theta$ with respect to the main probe beam  (Fig. (\ref{fig1})).  Note that AB does not hit the detector, but the scattered field of this beam may provide additional contribution to the detected polarimetric signal, and our goal now  is to calculate value of this contribution.     
 
 The calculation can be performed in the same way as in the previous section  with the following changes. 
The scattered field is calculated using Eq. (\ref{a7}) with the field ${\bf E_0(r)}$ replaced by $ {\bf E_0^t(r)}$, where $ {\bf E_0^t(r)}$ represents the field of the auxiliary (tilted)  beam. The field $ {\bf E_0^t(r)}$ can be obtained by  rotating ${\bf E_0(r)}$ by the angle $\Theta$ around the axis $(\cos\phi,\sin\phi,0)$ parallel to the  direction of polarization of the probe beam \footnote{Thus,  polarizations of the tilted  and the probe beams are the same}:
\begin{equation}
 {\bf E_0^t(r)}=M {\bf E}_0(M{\bf r}).
 \label{a20}
  \end{equation}
Here, the matrix $M$ is defined as

 \begin{equation}
  M=R(-\phi)H(\Theta)R(\phi)=\left (\matrix {\cos\Theta\sin^2\phi +\cos^2\phi&[1-\cos\Theta]\sin \phi\cos\phi & -\sin\phi\sin\Theta \cr [1-\cos\Theta]\sin\phi\cos\phi& \cos\Theta\cos^2\phi+\sin^2\phi &\cos\phi\sin\Theta \cr \sin\Theta\sin\phi &-\sin\Theta\cos\phi & \cos\Theta}\right )=
  \label{a21}
  \end{equation}
  $$
  =\left (\matrix {1-{1\over 2}\Theta^2\sin^2\phi &{1\over 2}\Theta^2\sin \phi\cos\phi & -\Theta\sin\phi \cr {1\over 2}\Theta^2\sin\phi\cos\phi & 1- {1\over 2}\Theta^2\cos^2\phi &\Theta\cos\phi \cr \Theta\sin\phi &-\Theta\cos\phi & 1-{1\over 2}\Theta^2}\right )+O(\Theta^3)
  $$
 Therefore, the field ${\bf E_0^t(r)}$ is defined by the expression
 \begin{equation}
                           {\bf E_0^t(r)}=  
          Q\sqrt{2 W_t\over c}
                           (\cos\phi,\sin\phi,0)\hskip1mm 
     \exp\imath\bigg [k Z({\bf r})-\omega t\bigg ]                                                 
                           \hskip1mm \exp\bigg [-{Q^2[X^2({\bf r})+Y^2({\bf r})]\over 4}\bigg ]
                           \label{a22}
                           \end{equation}
where 

 \begin{equation}
  \left (\matrix {X ({\bf r})\cr Y({\bf r})\cr Z({\bf r})}\right )\equiv
  \left (\matrix {\cos\Theta\sin^2\phi +\cos^2\phi&[1-\cos\Theta]\sin \phi\cos\phi & -\sin\phi\sin\Theta \cr [1-\cos\Theta]\sin\phi\cos\phi& \cos\Theta\cos^2\phi+\sin^2\phi &\cos\phi\sin\Theta \cr \sin\Theta\sin\phi &-\sin\Theta\cos\phi & \cos\Theta}\right )\left (\matrix {x \cr y\cr z}\right ) +\left (\matrix {\delta x \cr \delta y \cr \delta z }\right )
  \label{a23}
  \end{equation}
with ${\bf r}=(x,y,z)$. We denote by $W_t$ intensity of the  AB and take into account its possible spatial shift $(\delta x,\delta y,\delta z)$. 
 Substituting ${\bf E_0^t(r’)}$  (\ref{a22})  into Eq. (\ref{a7}) instead of ${\bf E_0(r’)}$, one can obtain   the following expression for the scattered field produced by AB:
                                     
   \begin{equation}
      \left (\matrix {{\cal E}_{1x}^t\cr {\cal E}_{1y}^t}\right ) =   
      \left (\matrix {-\sin\phi \cr \cos\phi }\right )
      \sqrt{2 W_t\over c}{Q k^2\over L}
       \int   
      \sin [k|{\bf r-r'}|+kZ'-\omega t]                           
                               \exp\bigg [-{Q^2(X'^2+Y'^2)\over 4}\bigg ]
       G_z({\bf r'}) d^3{\bf r'}
      \label{a24}
      \end{equation}
      where $X'=X({\bf r'})$, $Y'=Y({\bf r'})$ and $Z'=Z({\bf r'})$, with the functions           $X({\bf r'})$,                       
     $Y({\bf r'})$, 
     $Z({\bf r'})$  defined by Eq. (\ref{a23}) with substitution $x,y,z\rightarrow x',y',z'$.
      This formula has the same sense as Eq. (\ref{a11}); for clarity we supply components of the scattered field by superscript $t$.
      Taking into account this replacements, one can get the relationship for polarimetric signal produced by the  AB (instead of Eq. (\ref{a12}))
                                       
       \begin{equation}
           u_1^t= -{4\sqrt{WW_t}k^3 \sin[2\phi]\over cL^2}                     
                            \int_{-l_x}^{l_x}dx\int_{-l_y}^{l_y}dy 
                                                 \exp\bigg [-{k^2(x^2+y^2)\over  Q^2L^2}\bigg ]
                                                 \times
                  \label {a25} 
       \end{equation}
       $$
       \times \int  
       \hskip1mm
       \cos \hskip1mm k\bigg [|{\bf r-r'}|+
       Z'-L-{x^2+y^2\over 2L}
       \bigg ]                         
             \hskip1mm \exp\bigg [-{Q^2(X'^2+Y'^2)\over 4}\bigg ]
                 G_z({\bf r'}) d^3{\bf r'}
       $$ 
Calculation of intergrals  can be made as in the previous section, and  the final result for the polarimetric signal produced by the  AB is:   
  \begin{equation}
                      u_1^t= -{4\sqrt{WW_t}k \pi Q^2\sin[2\phi]\over c}                     
                   \int_V   \cos k[z'-Z']
                       \exp\bigg [-{Q^2(x'^2+y'^2+X'^2+Y'^2)\over 4}\bigg ]
                                G_z({\bf r'}) d^3{\bf r'}                   
                             \label {a26}  
                 \end{equation}           
                 where                        ${\bf r'}=(x',  y',  z')$
          \begin{equation}
          \left (\matrix {X' \cr Y' \cr Z'}\right )=
          \left (\matrix {\cos\Theta\sin^2\phi +\cos^2\phi&[1-\cos\Theta]\sin \phi\cos\phi & -\sin\phi\sin\Theta \cr [1-\cos\Theta]\sin\phi\cos\phi& \cos\Theta\cos^2\phi+\sin^2\phi &\cos\phi\sin\Theta \cr \sin\Theta\sin\phi &-\sin\Theta\cos\phi & \cos\Theta}\right )\left (\matrix {x' \cr y' \cr z'}\right ) +\left (\matrix {\delta x \cr \delta y \cr \delta z }\right )
          \label{a27}
          \end{equation}   
          One can see that  $u_1^t$ is proportional to overlap of the two beams and vanishes at large shifts $\delta x,\delta y,\delta z$.   The trigonometric factor $\cos k[z’-Z’]$, in fact, singles out harmonic of the gyrotropy with the spatial frequency equal to  difference between the wave vectors of  the two beams.                    
          Total signal in the presence of two beams  is the sum of  (\ref{a18})  and (\ref{a26}): $u_1+u_1^t$.   Remind that the angle $\Theta$ should not be  too large; otherwise, one should take into account the component ${\bf E_1^2(r)}$ in Eq. (\ref{a7}).

        \section{ Noise signal in the two-beam configuration } 
           
          The noise signal produced   by   the two beams in the configuration of Fig. 1 is calculated as Fourier transform of correlation function of  the total polarimetric signal  $u=u_1+u^t_1$. It  consists of 3 terms: 
\begin{equation}
{\cal N}_t(\nu)= \int dt e^{\imath \nu t}\langle u(0)u(t)\rangle=\int dt e^{\imath \nu t}\bigg [\langle u_1(0)u_1(t)\rangle+2\langle u_1(0)u_1^t(t)\rangle+\langle u_1^t(0)u_1^t(t)\rangle\bigg ]
\label{a28}
 \end{equation}
Using Eqs. (\ref{a18})  and (\ref{a26}), one can write the  expressions for each of them. The first term has been already calculated and is given by Eq. (\ref{a19}). For the correlator entering the last term, we have   

   \begin{equation}
      \langle u_1^t(0)u_1^t(t)\rangle=
      {16 WW_tk^2\pi^2Q^4\sin^2[2\phi]\over c^2}     
            \int_V d^3{\bf r}\int_V d^3{\bf r'}
            \cos k[z-Z]\cos k[z'-Z']\times
      \label{a29}
      \end{equation}
      $$
      \times\exp\bigg [-{Q^2(X^2+Y^2+x^2+y^2+X'^2+Y'^2+x'^2+y'^2)\over 4}\bigg ]\times
                                               \langle G_z({\bf r'},0)G_z({\bf r},t)\rangle,
 $$         
       
    where  $x,y,z\rightarrow {\bf r}$  and $X,Y,Z$ are defined by Eq. (\ref{a27})  
    
            \begin{equation}
            \left (\matrix {X \cr Y \cr Z}\right )=
            \left (\matrix {\cos\Theta\sin^2\phi +\cos^2\phi&[1-\cos\Theta]\sin \phi\cos\phi & -\sin\phi\sin\Theta \cr [1-\cos\Theta]\sin\phi\cos\phi& \cos\Theta\cos^2\phi+\sin^2\phi &\cos\phi\sin\Theta \cr \sin\Theta\sin\phi &-\sin\Theta\cos\phi & \cos\Theta}\right )\left (\matrix {x \cr y \cr z}\right ) +\left (\matrix {\delta x \cr \delta y \cr \delta z }\right )
            \label{a30}
            \end{equation} 
         $X',Y',Z'$ are  similar  functions of  $x',y',z'\rightarrow {\bf r'}$.
      
      Finally, the cross correlator $\langle u_1^t(0)u_1(t)\rangle$ can be written as
     \begin{equation}
               \langle u_1^t(0)u_1(t)\rangle=
               {16 W\sqrt{WW_t}k^2\pi^2Q^4\sin^2[2\phi]\over c^2}     
                     \int_V d^3{\bf r}\int_V d^3{\bf r'}
                     \cos k[z-Z]\times
               \label{a31}
               \end{equation}
               $$
               \times\exp\bigg [-{Q^2(X^2+Y^2+x^2+y^2)\over 4}-{Q^2(x'^2+y'^2)\over 2}\bigg ]\times
                                                        \langle G_z({\bf r'},0)G_z({\bf r},t)\rangle
          $$
    Consider now physical sense of different  factors entering     Eqs.(\ref{a19}), (\ref{a29}), and (\ref{a31}).

{\bf Exponential factor }
  reduces the region of integration down to the region of overlapping of  the two beams. If $\Theta$ is not too large and $l_s\Theta < w$, this region is close to “the beam volume within the sample”.  In this case, the exponential factor can be calculated at
 $X=x,Y=y,Z=z,X'=x',Y'=y',Z'=z'$. Note that  it is rather difficult  to satisfy the condition $ l_s \Theta < w$   in a real experiment. For this reason,   the overlapping factor may  considerably reduce contribution of the  AB to the polarimetric signal.     
    
 {\bf Trigonometric factor } at small angles $\Theta$ is controlled by the difference between wave vectors of the two beams because the cosine argument can be evaluated  as     $z-Z=[\cos\phi y-\sin\phi x]\Theta$. 
  
  {\bf Correlation function } $\langle G_z({\bf r'},0)G_z({\bf r},t)\rangle$ is determined by particular  
model of the gyrotropic medium. For homogeneous media, it depends on  the difference ${\bf r-r'}$ 
of the spatial arguments. For the model of independent spins, described below
$\langle G_z({\bf r'},0)G_z({\bf r},t)\rangle$ $\sim \delta ({\bf r-r'})e^{-|t|/\tau}\cos\omega_0t$

Thus, the integrals entering Eqs.(\ref{a19}), (\ref{a29}), and (\ref{a31}) can
 be calculated for any particular model of the gyrotropic medium. In the next section,  we will present calculations for  the model 
of independent paramagnetic particles (spins). Still, the following general remark should be made.
Let the  beam waist $4w$ and the sample length $l_s$ be much greater than   
the gyrotropy correlation radius $R_c$ and  spatial period $2\pi/k\Theta$  related to 
 the difference of wave vectors of the two beams: $4w,l_s\gg R_c,2\pi/k\Theta$.
 Then, one can substitute variables in the integrals entering Eqs.(\ref{a19}), (\ref{a29}), and (\ref{a31}) 
in the following way: ${\bf r,r'}\rightarrow {\bf R\equiv  r-r',R'\equiv r+r'}$ 
 and take advantage of the fact that the correlator 
  $ \langle G_z({\bf r'},0)G_z({\bf r},t)\rangle$
depends on  difference of its arguments:
\begin{equation}
  \langle G_z({\bf r'},0)G_z({\bf r},t)\rangle\equiv K({\bf r-r'},t)
\label{a32}
\end{equation}

Then,  the integral over  ${\bf R\equiv r-r'}$  in Eq. (\ref{a19}) can be estimated as  
the average of $K({\bf R},t)$ over  irradiated volume of the sample $V_b $. The integration over
 ${\bf R'\equiv r+r'}$ gives this  volume itself, and we obtain    

\begin{equation}
{\cal N}(\nu)={16 W^2k^2\pi^2Q^4\sin^2[2\phi]\over c^2}
\int dt\hskip1mm  e^{\imath \nu t}\int_V d{\bf r} d{\bf r'}
\exp\bigg [-{Q^2(x'^2+y'^2+x^2+y^2)\over 2}\bigg ]K({\bf r-r'},t)\sim
\label{a33}
\end{equation}
$$
\sim {W^2l_s\sin^2[2\phi]\over S}\int dt\hskip1mm e^{\imath\nu t}\int_{V_b} d{\bf R} \hskip1mm K({\bf R},t).
  $$                  
        Here, we denote the cross section area of the beam by $S\equiv 4\pi w^2$ and take into account that $w=1/Q$ and 
 that irradiated  volume of the sample is 
 $V_b= Sl_s$, where $l_s$ 
 is the sample length. We come to the known result  that the noise power signal is proportional to the sample length and inversely proportional to the beam cross section \cite{Zap, S,S1}. 

The correlation function Eq.(\ref{a29}) can be estimated in a similar way.  
If $\Theta$ is not too large, then the arguments of the cosine functions can be evaluated as   
$z-Z=[\cos\phi y-\sin\phi x]\Theta$ and $z'-Z'=[\cos\phi y'-\sin\phi x']\Theta$.
  Therefore,  one can represent the  product of the cosine functions in  Eq.(\ref{a29}) as
    $$
    \cos k[z-Z]\cos k[z'-Z']={1\over 2}\cos \bigg\{k\Theta \bigg [(y-y')\cos\phi -(x- x')\sin\phi \bigg ] \bigg\} +
    $$
$$
+{1\over 2}\cos \bigg\{k\Theta \bigg [(y+y')\cos\phi -(x+x')\sin\phi \bigg ] \bigg\}
$$

Note that the difference $\Delta {\bf k}$ between the wave vector of the two beams  
for small $\Theta$ has only $x$ and $y$ components: $\Delta {\bf k}=k\Theta (-\sin\phi,\cos\phi,0)$.
Therefore,  this relationship after substitution of variables  
${\bf r,r'}\rightarrow {\bf R= r-r',R'=r+r'}$ takes the form
    $$
    \cos k[z-Z]\cos k[z'-Z']={1\over 2}\cos (\Delta {\bf k,R}) +
{1\over 2}\cos (\Delta {\bf k,R'})
    $$

Remind that our treatment is valid when  
 $w$ is large enough ($\Delta kw>2\pi$). In this case, 
 the integral $\int_{V_b}d{\bf R'}\cos (\Delta {\bf k,R'})\sim 0$,
 and we come to conclusion that the  
  correlation function Eq. (\ref{a29}) can be estimated as follows 

   \begin{equation}
      \langle u_1^t(0)u_1^t(t)\rangle\sim
      {WW_t Q^4\sin^2[2\phi]}     
            \int_{V_b} d{\bf R}d{\bf R'}\hskip1mm K({\bf R},t)
\cos (\Delta {\bf k,R})  \sim
      \label{a34}
      \end{equation}
  $$
\sim  {WW_t \sin^2[2\phi]l_s\over S}     
            \int_{V_b} d{\bf R}  \hskip1mm K({\bf R},t) \cos (\Delta {\bf k,R})         
$$

 Thus, contribution of the auxiliary tilted beam (AB) to  the noise signal is proportional to Fourier 
transform of  the correlation function of gyrotropy at spatial frequency equal to difference of the wave vectors of the two beams ($\Delta {\bf k}$).

Therefore, by measuring dependence of the noise signal, in the two-beam configuration, on the angle between the beams (in fact, on $\Delta k$) and using the inverse Fourier transform, one can restore spatial dependence of the gyrotropy correlation function $K({\bf R},t)$. Recall that in the conventional spin noise spectroscopy, only temporal dependence of this correlation function averaged over the irradiated volume of the sample is revealed. 

Similarly,  it can be shown that, under these conditions, contribution of  the cross correlator 
Eq. (\ref{a31}) is relatively small.

   \section { The model of independent spins }
In this model, the random field of gyrotropy $G_z({\bf r})$ has the form
    \begin{equation}
      G_z({\bf r})=\sum_{i=1}^N g_i(t)\delta ({\bf r-r}_i),
      \label{a35}
      \end{equation}
thus corresponding  to $N$ paramagnetic particles (spins)  randomly distributed over the volume of the medium with
  some average density $\sigma\equiv N/V$, where $V$ is the total  volume of the system.
  We assume that $g_i(t)$ is proportional to $z$-component of magnetization of the $i$-th particle.     The polarimetric signal can be calculated using Eq.(\ref{a18}):

 \begin{equation}
                              u_1=u_1(t)= -{4Wk \pi Q^2\sin[2\phi]\over c}                     
                           \int_V   
                               \exp\bigg [-{Q^2(x'^2+y'^2)\over 2}\bigg ]
                                         \sum_i g_i(t)\delta ({\bf r'-r_i})d^3{\bf r'}                   
                                     \label {a36} 
                          \end{equation}
 Let us calculate  polarimetric signal $u_{10}$ for the sample in which all
 magnetizations $g_i(t)$ are constant and the same: $g_i(t)=g_0=$ const.  
This corresponds to a paramagnet  in a high magnetic field at low temperature.  In this case, Eq. (\ref{a36}) gives 

      \begin{equation}
  u_{10}= -{8Wk g_0\sigma l_s\pi^2 \sin[2\phi]\over c}                     
          \label {a37} 
    \end{equation}    
 We will see below that the quantity $u_{10}$ provides us a convenient scale. 
 Let us now consider  the gyrotropic medium with the quantities $g_i$ changing randomly in a stationary way with the correlation function  $\langle g_i(t)g_k(t')\rangle =\delta_{ik}K(t-t’)$ (should be distinguished from the spatiotemporal correlation function of Eq. (\ref{a32})) and calculate, for this model,  the noise power spectrum using Eq. (\ref{a19}). We have
 \begin{equation}
  \langle G_z({\bf r'},0)G_z({\bf r},t)\rangle={1\over V}\sum_{i}\int d^3{\bf r_i}
  \langle g_i(0)g_i(t)\rangle \delta ({\bf r'-r_i}) \delta ({\bf r-r_i})=\delta ({\bf r-r'})\sigma K(t),
  \label{a38}
  \end{equation}     
and, consequently, 
    \begin{equation}
          {\cal N}(\nu)={16 W^2k^2\pi^3Q^2l_s\sigma\sin^2[2\phi]\over c^2}     
          \int dt\hskip1mm  e^{\imath \nu t} K(t)                 
\label{a39}          
\end{equation}

If we accept for  beam area the expression $S=4\pi w^2$, then $Q^2=1/w^2=4\pi/S$.  Taking  into account Eq. (\ref{a37}),   we obtain the expression for noise power spectrum

\begin{equation}
{\cal N}(\nu)={u_{10}^2\over  \sigma l_sS}\hskip1mm \int dt e ^{\imath \nu t}{\langle g(0)g(t)\rangle\over g_0^2}
\label{a40}
\end{equation}     

Note that   $\sigma  l_s S\equiv N_b$ is the number of spins in the irradiated  volume of the sample.

 In the simplest case, each  paramagnetic particle  of the gyrotropic medium can be associated with the effective  spin 1/2. Then, the total magnetization can be expressed as:  $g_0^2=(g\beta)^2 /4$ (here, $g$ is the effective  $g$-factor and $\beta$ is the Bohr magneton). In the presence of the transverse magnetic field $B_x$, the correlator $\langle g(0)g(t)\rangle$ can be calculated using the following chain of relationships:
\begin{equation}
\langle g(0)g(t)\rangle ={(g\beta)^2 \over 2}\hbox{ Sp }[S_zS_z(t)+S_z(t)S_z]\rho_{eq}\hskip 10mm
S_z(t)=e^{-\imath \omega_0tS_x}S_ze^{\imath \omega_0tS_x} \hskip5mm \omega_0\equiv {g\beta B_x\over\hbar}
\label{a41} 
\end{equation}    
Here, $\rho_{eq}$ is the density matrix of the  two-level system representing our effective spin 1/2.
 If the temperature is high enough ($kT\gg g\beta B_x$), the density matrix can be taken constant,
 $\rho_{eq}=\hat I/2$ ($\hat I$  is the unit matrix),  and we obtain 
\begin{equation}
\langle g(0)g(t)\rangle = {(g\beta)^2\over 4}\hbox{ Sp }[S_zS_z(t)+S_z(t)S_z] 
={(g\beta)^2\over 4}\hskip1mm \cos\omega_0 t \rightarrow 
{(g\beta)^2\over 4}\hskip1mm e^{-|t|/\tau}\cos\omega_0 t \hskip5mm \Rightarrow 
\label{a42}
\end{equation} 
$$
\Rightarrow {\langle g(0)g(t)\rangle\over g_0^2}=e^{-|t|/\tau}\cos\omega_0 t
$$
 We introduce phenomenologically the transverse relaxation time $\tau$. 
 So, for the noise power spectrum we have
\begin{equation}
{\cal N}(\nu)={u_{10}^2\tau\over N_b}\hskip1mm \bigg [
{1\over 1+(\omega_0+\nu)^2\tau^2} + {1\over 1+(\omega_0-\nu)^2\tau^2}
\bigg ]
\label{a43}
\end{equation}      
The root-mean-square value of the polarimetric noise is given by the relationship 
 \begin{equation}
 \langle \delta u^2\rangle={1\over 2\pi}\int  {\cal N}(\nu)d\nu=  {u_{10}^2\over  N_b}
\label{a44}
\end{equation} 
 In a similar way, one can calculate the power spectrum of the polarimetric noise in the presence of the auxiliary beam AB. Using Eq. (\ref{a37}) for the correlation function and Eqs. (\ref{a19}), (\ref{a29}), and (\ref{a31}), we obtain:  

\begin{equation}
             \langle u(0)u(t)\rangle=
             u_{10}^2{e^{-|t|/\tau}\cos\omega_0t \over N_b}\bigg [
             1+2\sqrt{W_t\over W}\exp\bigg [-{k^2\Theta^2\over 4Q^2}\bigg ]+ {W_t\over W}{1\over 2}\bigg (
                          1+\exp \bigg [-{k^2\Theta^2\over Q^2}\bigg ]\bigg )
             \bigg ]
               \label{a45}
             \end{equation}
     
 If $2\pi/k=1\ \mu$m, $\Theta\sim 0.1$, and $1/Q\sim 30\ \mu$m, the exponential factors can be omitted, and  simplified expressions for the correlation function and the noise power spectrum acquire the form
 
 \begin{equation}
              \langle u(0)u(t)\rangle= 
              {u_{10}^2 \over N_b}\bigg [
              1+ {W_t\over 2W}\bigg ]e^{-|t|/\tau}\cos\omega_0t
                \label{a46}
              \end{equation}

\begin{equation}
{\cal N}(\nu)={u_{10}^2\tau\over N_b}\hskip1mm \bigg [
              1+ {W_t\over 2W}\bigg ] \bigg [
{1\over 1+(\omega_0+\nu)^2\tau^2} + {1\over 1+(\omega_0-\nu)^2\tau^2}
\bigg ]
\label{a47}
\end{equation} 
It is seen from Eq. (\ref{a47}) that if $W_t\sim W$, then switching the axiliary beam on leads to 50$\%$ increase of the noise power  and, therefore,  can  be    easily observed.
Note once again that we assumed complete overlapping of the two beams. Therefore, the contribution of the auxiliary beam  to the noise power spectrum in real experiments, when this is not the case, may be somewhat smaller.    

The above treatment was performed for the case of absence of any spatial correlation in the field of gyrotropy. The rfesult of this assumption is the absence of any dependence of the noise signal on the angle  $\Theta$ (at small $\Theta$). In the presence of spatial correlation of the gyrotropy, the noise signal will decrease with $\Theta$ (with increasing $\Delta k=k\Theta$). Specifically, if the noise signal decreases, say, by a factor of 2 at an angle of $\Theta_{1/2}$, then the correlation radius of the gyration field $R_c$ can be estimated as $R_c\sim [\Delta k_{1/2}]^{-1}\equiv [k \Theta_{1/2}]^{-1}$.

\section*{Conclusion}

The main goal of the paper was to understand deeper the role and properties of the scattered field underlying signal formation in the Faraday-rotation-based spin noise spectroscopy. The sample with fluctuating spins is considered as an optical medium with its gyrotropy fluctuating both in time and in space. The noise signal arising due to heterodyning of the light scattering by the inhomogeneous medium is calculated for a focused Gaussian beam in the single-scattering approximation.  We show that, in real experiments, only a small fraction of the scattered field contributes to the detected signal, namely, onle components of the scattered field that overlap with those of the probe beam in the momentum space. Therefore, a more efficient use of the scattered field, in spin noise spectroscopy, can be achieved by increasing this overlap in proper optical arrangements. Our calculations  confirm the common assumption that the noise signal, in the conventional geometry of spin noise spectroscopy, is proportional to the sample's gyrotropy spatially averaged over the irradiated volume. We also consider a two-beam experimental arrangement in which properties of the scattered light field are revealed in a much more pronounced way. We show that the signal produced by the auxiliary  light beam tilted with respect to the probe is proportional to Fourier transform of the gyrotropy at spatial frequency equal to difference of the wave vectors of two beams.  Accordingly, in the presence of spatial correlation of the gyrotropy field, Fourier components at higher spatial frequencies will appear to be suppressed, and  contribution of the auxiliary beam into the noise signal at larger angles between the beams will decrease. This effect can be used to investigate spatial correlation of spins in spin noise spectroscopy. The results of rigorous solution of the problem are presented here for the case of spatially uncorrelated gyrotropy with, correspondingly, "white" spatial spectrum of the gyrotropy fluctuations. 
The results of this study are important from the viewpoint of fundamental physics of signal formation in the spin noise spectroscopy and , at the same time, may be useful for increasing sensitivity of the spin-noise technique.

\newpage
%\begin{figure}
%\epsffile{fig1.ps}
%\caption{(a) -- зависимость критерия Андерсона $D$ от параметра беспорядка $\Delta$
% для модели Ллойда. Гладкая линия -- расчет по формулам (\ref{4}), зашумленная линия -- компьютерный эксперимент.
%(b) -- распределение степени локализации состояний $W(U)dU$ для одномерной  модели Ллойда при
% $\Delta=0.007$ и $dU=7/300$. Гладкая кривая -- расчет по формуле (\ref{4}), зашумленная кривая -- компьютерный
% эксперимент, при котором было выполнено усреднение по 2000 реализаций случайных матриц (\ref{1})
%   размером $N=2000$.  }
%\end{figure}

%\begin{figure}
%\epsffile{fig2.ps}
%\caption{Пояснения в тексте.}
%\end{figure}

\end{document}